\documentclass{ws-procs975x65}

\def\beq{\begin{equation}}
\def\eeq{\end{equation}}

\begin{document}

\title{Removal of angular momentum by strong magnetic field stresses in advective accretion flows around black holes}
\author{Banibrata Mukhopadhyay}

\address{Department of Physics, Indian Institute of Science, Bangalore 560012\\
E-mail: bm@physics.iisc.ernet.in\\
}

\author{Koushik Chatterjee} 

\address{Department of Physics, Indian Institute of Technology Kharagpur, Kharagpur 721302\\
E-mail: kchatterjee009@gmail.com\\
}

\begin{abstract}
We show that the removal of angular momentum is possible in the presence of large scale magnetic stresses, 
arisen by fields much stronger than that required for magnetorotational instability, in geometrically thick, advective, 
sub-Keplerian accretion flows around black holes in steady-state, in the complete absence of alpha-viscosity.
The efficiency of such angular momentum transfer via Maxwell stress, with the field 
well below its equipartition value, could be equivalent to that of 
alpha-viscosity, arisen via Reynolds stress, with $\alpha=0.01-0.08$. 
We find in our simpler vertically averaged advective disk model that stronger the magnetic field and/or larger the vertical-gradient of 
azimuthal component of magnetic field, stronger the rate of angular momentum transfer is, which in turn may 
lead to a faster rate of outflowing matter, which has important implications to describe the hard spectral 
states of black hole sources. When the generic origin of alpha-viscosity is still being explored,
mechanism of efficient angular momentum transfer via magnetic stresses alone is very interesting.
\end{abstract}

\bodymatter


\section{Introduction}\label{intro}

Blandford and Payne \cite{blandpayne} showed angular momentum transfer via large magnetic stresses, 
in the absence of alpha-viscosity, in the framework of self-similar
Keplerian disk flows. Here, we show in a simpler 1.5-dimensional, vertically averaged
disk model that the Maxwell stresses due to strong magnetic fields are adequate enough 
for angular momentum transfer even in advective accretion flows, without self-similar assumption, 
describing the hard spectral states of black hole sources. 

The idea of exploring magnetic stress in order to explain astrophysical systems is not
really new. This was implemented, for example, in the solar wind which was understood to 
have decreased Sun's angular momentum through the effect
of magnetic stresses (see, e.g., Ref.~\refcite{waber}), in the proto-stellar gas clouds 
which might have been contracted by magnetic effects \cite{Mouschovias}. 
Ozernoy and Usov \cite{usov} and Blandford \cite{bland} showed that the energy is possible to 
extract continuously by electromagnetic torques and
twisted field lines in accretion disks. By linear stability analysis of the accretion disks,
it was shown by Cao and Spruit \cite{cao} that angular momentum is possible to remove by the 
magnetic torque exerted by a centrifugally driven wind. 
However, by solving the local vertical structure
of a geometrically thin accretion disk threaded by a poloidal magnetic field, 
Ogilvie and Livio \cite{ogil} showed the shortcoming of launching an outflow and suggested
for an existence of additional source of energy for its successful launching.

Here we demonstrate, semi-analytically, the effects of strong magnetic field, stronger than
that needed for magnetorotational instability (MRI) \cite{sujit}, with 
plasma$-\beta > 1$ yet, on to the vertically averaged advective accretion flows in 
vertical equilibrium in order to transport matter. 
Therefore, we consider the flow variables to depend on the radial coordinate only. 
Although, in reality, a non-zero vertical magnetic field should induce a vertical motion,
in the platform of the present assumption, any vertical motion will be featured as an outward motion. 
Indeed, our aim here is to furnish removal of angular momentum from the flow via
magnetic stresses, independent of its vertical or outward transport.

\section{Basic model equations}\label{model}

We describe optically thin, magnetized, viscous, axisymmetric, advective, vertically averaged, steady-state accretion
flow, in the pseudo-Newtonian framework with the Mukhopadhyay \cite{m02} potential. Hence, the equation 
of continuity, vertically averaged hydromagnetic equations for energy-momentum balance in different directions are 
given by (assuming that the dimensionless variables do not vary significantly in the vertical direction such that 
$\partial/ \partial z \sim s_i/h$ and, as a consequence, the vertical component of velocity is zero),
\begin{eqnarray}
\nonumber
&&\dot{M}=4\pi  x \rho h \vartheta,~~~ 
\vartheta\frac{d\vartheta}{dx}+\frac{1}{\rho}\frac{dP}{dx}-\frac{\lambda^{2}}{x^3}+F=\frac{1}{4\pi \rho}\left(B_x\frac{dB_x}{dx}+s_1\frac{B_xB_z}{h}-\frac{B_{\phi}^2}{x}\right),\\
\nonumber
&&\vartheta\frac{d\lambda}{dx}=\frac{1}{x\rho}\frac{d}{dx}\left(x^2W_{x\phi}\right)+\frac{x}{4\pi \rho}\left(B_x\frac{dB_{\phi}}{dx}+s_2\frac{B_zB_{\phi}}{h}+\frac{B_xB_{\phi}}{x}\right),\\
&&\frac{P}{\rho h}=\frac{Fh}{x}-\frac{1}{4\pi \rho}\left(B_x\frac{dB_z}{dx}+s_3\frac{B_z^2}{h}\right),\\
\nonumber
&&\vartheta T\frac{ds}{dx}=\frac{\vartheta}{\Gamma_3-1}\left(\frac{dP}{dx}-\frac{\Gamma_1P}{\rho}\frac{d\rho}{dx}\right)
=Q^+-Q^-=Q^+_{vis}+Q^+_{mag}-Q^-_{vis}-Q^-_{mag},
\end{eqnarray}
where $W_{x\phi}=\alpha(P+\rho \vartheta^2)$ and $\alpha$ the Shakura-Sunyaev viscosity parameter \cite{ss}.
Note that $s_1$, $s_2$ and $s_3$ are the degrees of vertical scaling for the radial, azimuthal and vertical components of the 
magnetic field respectively. 
Here $\dot{M}$ is the 
conserved mass accretion rate, $\rho$ the mass density of the flow, $\vartheta$ the radial velocity, $P$ the total 
pressure including the magnetic contribution, $F$ the force corresponding to the pseudo-Newtonian potential for 
rotating black holes \cite{m02}, $\lambda$ the angular momentum per unit mass, $W_{x\phi}$ the viscous shearing stress written 
following the Shakura-Sunyaev prescription \cite{ss} with appropriate modification \cite{mg03},
$h \sim z$, the half-thickness and $x$ the radial coordinate of the disk, when both of 
them are expressed in units of $GM/c^2$, where $G$ the Newton's gravitation constant, $M$ the mass of black hole, 
$c$ the speed of light, $s$ the entropy per unit volume, 
$T$ the (ion) temperature of the flow, $Q^+$ and $Q^-$ are the net rates of energy released and radiated out per unit 
volume in/from the flow respectively (when $Q^+_{vis}$, $Q^+_{mag}$, $Q^-_{vis}$, $Q^-_{mag}$ are the respective 
contributions from viscous and magnetic parts). All the variables are made dimensionless in the spirit of 
dimensionless $x$ and $z$. We further assume, for the present purpose, the heat radiated 
out proportional to the released rate with the proportionality constants $(1 - f_{vis})$ and $(1 - f_m)$, respectively, 
for viscous and magnetic parts of the radiations. $\Gamma_1$, $\Gamma_3$, which are functions of polytropic constant $\gamma$, 
indicate the polytropic indices depending 
on the gas and radiation content in the flow (see, e.g., Ref.~\refcite{rm1}, for exact expressions) 
and $B_x$, $B_{\phi}$ and $B_z$ are the components of magnetic field. The model for $Q^+_{vis}$ is taken from the 
previous work \cite{rm1}, and the relation for $Q^+_{mag}$ is taken from that by Bisnovatyi-Kogan and 
Ruzmaikin \cite{bis}.

Hydromagnetic flow equations must be supplemented by (for the present purpose, steady-state) equations of induction
and no magnetic monopole at the limit of very large Reynolds number, given by
\begin{eqnarray}
\nabla \times \vec{v} \times \vec{B}=0,~~~
\frac{d}{dx}(xB_x)+s_3\frac{B_z}{h}=0,
\end{eqnarray}
when $\vec{v}$ and $\vec{B}$ are respectively the velocity and magnetic field vectors and $\nu_m$ is the magnetic 
diffusivity. 

\section{Solutions and Results}\label{sol}

We take into account two situations. (1) Flows with a relatively higher $\dot{M}$ and, hence, lower $\gamma$, 
modelled around stellar mass black holes: such flows may or may not form Keplerian accretion disks. (2) Flows with 
a lower $\dot{M}$ and, hence, higher $\gamma$, modelled around supermassive black holes: such flows are necessarily 
hot gas dominated advective (or advection dominated) accretion flows.

We find that the flows with plasma-$\beta > 1$, but $\alpha=0$, exhibit adequate matter transport, as efficient as the 
$\alpha$-viscosity with $\alpha= 0.08$, but without magnetic stresses, would do. This is interesting as the origin of $\alpha$ (and the corresponding 
instability and turbulence) is
itself not well understood. The maximum required large scale magnetic field is $\sim 10^5$G in a disk
around $10M_{\odot}$ black holes and $\sim10$G in a disk around $10^7M_{\odot}$ supermassive black holes,
where $M_\odot$ is solar mass. 
The presence of such a field, in particular for a stellar mass black hole disk when the binary companion supplying 
mass is a Sun-like star with
the magnetic field on average 1G, may be understood, if the field is approximately frozen with the disk fluids (or the
supplied fluids from the companion star remain approximately frozen with the magnetic field) or disk fluids exhibit very
large Reynolds number. Indeed, all the present computations are done at the limit of large Reynolds number, as really
is the case in accretion flows, such that the term associated with the magnetic diffusivity in the induction equation is
neglected. The size of a disk around supermassive black holes is proportionately larger compared to that around
a stellar mass black hole. Hence, from the equipartition theory, indeed the magnetic field is expected to be decreased
here compared to that around stellar mass black holes. Figure 1 shows a typical set of accretion solutions and confirms that
flows around a stellar mass black hole with $\alpha$-viscosity but without large scale magnetic fields (viscous flow)
show similar behavior to that with large scale magnetic fields 
without $\alpha$-viscosity (magnetic flow). The flows around a supermassive black hole show very similar features, except
with reduced field strength. For other details, see Ref.~\refcite{kou}.

\begin{figure}[h]
\begin{center}
\includegraphics[angle=0,width=2.5in]{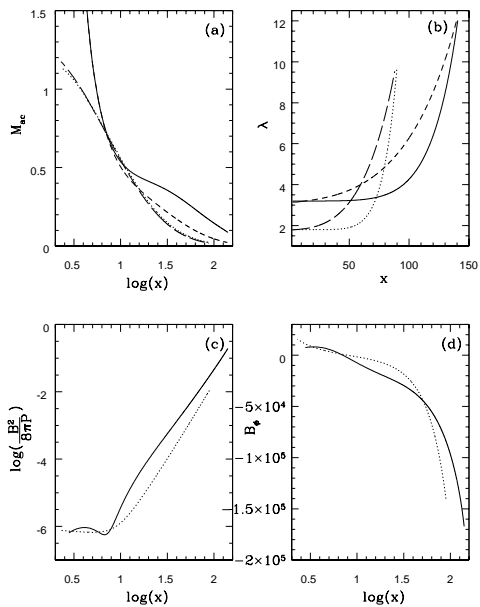}
\end{center}
\caption{
(a) Mach number, (b) angular momentum per unit mass in $GM/c$, (c) inverse of plasma-$\beta$,
(d) azimuthal component of magnetic field in G,
when solid and dotted lines are for magnetic flows around
Schwarzschild ($a=0, \lambda_c=3.2$) and Kerr ($a=0.998, \lambda_c=1.8$) black holes respectively,
and dashed and long-dashed lines are for viscous flows
around Schwarzschild ($a=0, \alpha=0.017, \lambda_c=3.15$) and Kerr ($a=0.998, \alpha=0.012, \lambda_c=1.8$)
black holes respectively, where $a$ is the dimensionless spin parameter of black hole and $\lambda_c$ the quantity at critical radius.
Other parameters are $M=10M_\odot$, $\dot{M}=0.1$ Eddington rate,
polytropic constant $\gamma=1.335, f_{vis}=f_m=0.5, s_2=-0.5$.
}
\label{stmasm}
\end{figure}

\begin{figure}[h]
\begin{center}
\includegraphics[angle=0,width=2.5in]{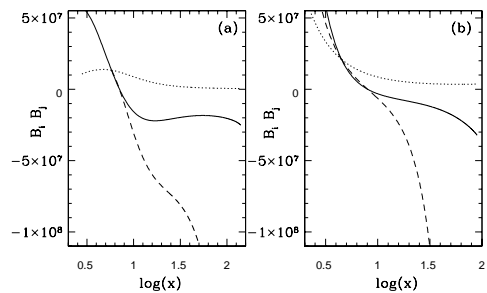}
\end{center}
\vskip-4cm
\caption{
Components of magnetic stress: $B_x B_\phi$ (solid line),
$B_x B_z$ (dotted line), $B_\phi B_z$ (dashed line), for (a) Schwarzschild
magnetic flow of Fig. \ref{stmasm}, (b) Kerr magnetic flow of Fig. \ref{stmasm}.
}
\label{stmasstr}
\end{figure}

Let us now explore in more details, how exactly various components of magnetic stress lead to
angular momentum transfer in the flows. 
Figure \ref{stmasstr}a shows that the stress component $B_x B_z$ around a Schwarzschild
black hole increases almost throughout
as matter advances towards the black hole. This implies that the flow is prone
to outflow through the field lines, which effectively helps in
infalling matter towards the black hole. However, in the near vicinity of black hole, $B_x B_z$ decreases,
as indeed outflow is not possible therein. 
In this flow zone, the angular momentum becomes very small which practically does
not affect the infall. The magnitude of $B_\phi B_z$ decreases till the
inner region of accretion flow, implying the part of matter to be spiralling out
and, hence, removing angular momentum leading to infall of matter.
Finally, the magnitude of $B_x B_\phi$
increases at a large and a small distances from the black hole (except around
the transition radius), which helps removing angular momentum and further infall. 
This is the same as the Shakura-Sunyaev viscous stress would do with the increase of matter pressure.
However, at the intermediate zone,
the angular momentum transfer through $B_x B_\phi$ reverses and a part of
the matter outflows. At the Keplerian to sub-Keplerian transition zone,
due to the increase of flow thickness, matter is vertically kicked effectively,
showing a decrease of $B_x B_\phi$. Majority of the features remain similar
for the flow around a rotating black hole, as shown in Fig. \ref{stmasstr}b.
However, a rotating black hole reveals a stronger/efficient outflow/jet in general. Hence, except at the inner zone,
$B_x B_\phi$ decreases throughout, which helps kicking the matter outwards by 
transferring the angular momentum inwards.

Different components of the magnetic stress tensor have different roles: $B_xB_{\phi}$ controls the infall in the disk plane,
whereas $B_{\phi}B_z$ renders the flow to spiral outwards and, hence, outflow. Moreover, $B_xB_z$ helps to kick the 
matter out vertically. Larger the field strength, larger is the power of magnetic stresses. Interestingly, 
the magnitude of magnetic
field decreases, as the steady-state matter advances towards the black hole. This is primarily because $B_{\phi}B_z$ 
(and also $B_xB_{\phi}$ for a rotating black hole) decreases inwards almost entirely in order to induce inflow
via angular momentum transfer through outflow. 
This further reveals
a decreasing $|B_{\phi}|$ as the output of self-consistent solutions of the coupled set of equations.

\section{Discussion and Conclusions}\label{last}

Is there any observational support for the existence of such a magnetic field, as required for the magnetic accretion
flows discussed here? Interestingly, the polarization measurements in the hard state of Cyg X-1 imply that it should
have at least 10mG field at the source of emission \cite{lau}. In order to explain such high polarization, a
jet model was suggested by Zdziarski et al. \cite{zd}, which requires a magnetic field $\sim (5 - 10) \times 10^5$G 
at the base of jet and hence in the underlying accretion disk. 

In the present computations, we have assumed the flow to be vertically averaged without allowing any vertical
component of the flow velocity. The most self-consistent approach,
in order to understand vertical transport of matter through the magnetic effects which in turn leads to the radial
infall of rest of the matter, is considering the flow to be moving in the vertical direction from the disk plane as well.
Such an attempt, in the absence of magnetic and viscous effects, was made earlier by one of the present authors \cite{deb} in the model
framework of coupled disk-outflow systems. In such a framework, the authors further showed that the outflow
power of the correlated disk-outflow systems increases with the increasing spin of black holes. Our future goal is now
to combine that model with the model of present work, so that the coupled disk-outflow systems can be investigated
more self-consistently and rigorously, when the magnetic field plays indispensable role in order to generate vertical
flux in the three-dimensional flows.

\section*{Acknowledgments}
K.C. thanks the Academies' Summer Research Fellowship Programme of India for offering him a Fellowship to
pursue his internship in Indian Institute of Science, Bangalore, 
when most of the calculations of this project were done.

\end{document}